\documentclass[a4paper,11pt]{article}
\usepackage{pos}
\usepackage{subfig}

\title{Optimizing HERON for 100 PeV Neutrino Detection}
\author*[a,b]{Andrew Zeolla}

\onbehalf{for the BEACON and GRAND Collaborations{\normalsize \\ \normalfont(a complete list of authors can be found at the end of the proceedings)}\\}

\affiliation[a]{Department of Physics, Pennsylvania State University, University Park, PA 16802, USA}
\affiliation[b]{Center for Multimessenger Astrophysics, Institute of Gravitation and the Cosmos, Pennsylvania State University, University Park, PA 16802, USA}
\emailAdd{azeolla@psu.edu}
\abstract{The Hybrid Elevated Radio Observatory for Neutrinos (HERON) is designed to target the astrophysical flux of Earth-skimming tau neutrinos at 100 PeV. HERON consists of multiple compact, phased radio arrays embedded within a larger sparse array of antennas, located on the side of a mountain. This hybrid design provides both excellent sensitivity and a sub-degree pointing resolution. To design HERON, a suite of simulations accounting for tau propagation, shower development, radio emission, and antenna response were used. These simulations were used to discover the array layout which provides maximum sensitivity at 100 PeV, as well to select the optimal antenna design. Additionally, the event reconstruction accuracy has been tested for various designs of the sparse array via simulated interferometry. Here, we present the HERON simulation procedure and its results.}

\ConferenceLogo{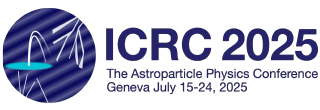}

\FullConference{39th International Cosmic Ray Conference (ICRC2025)\\
 15–24 July 2025\\
Geneva, Switzerland\\}
\begin{document}
\maketitle

\section{Introduction}
Astrophysical neutrinos point back to their source and can propagate for extreme distances, allowing us to study the most energetic objects in the universe \cite{Ackermann:2019ows}. This has been demonstrated by IceCube which has detected astrophysical neutrinos with energies up to 100 PeV, and identified two potential neutrino sources \cite{IceCube:2018cha, IceCube:2022der}. Recently, KM3NeT has detected an ultrahigh energy (UHE) neutrino \cite{KM3NeT:2025npi}. UHE tau neutrinos in particular present a unique possibility for detection. UHE tau neutrinos which skim the Earth can interact within, producing a tau lepton, which then escapes into the lower atmosphere before immediately decaying \cite{Zas_2005}. This decay initiates an up-going extensive air shower (EAS), which in turn emits a brief radio impulse. Two planned experiments which seek to detect this radio emission are GRAND and BEACON.

The Giant Radio Array for Neutrino Detection (GRAND) concept consists of multiple large, sparse radio antenna arrays \cite{GRAND, Olivier}. Each sparse array consists of 10,000 autonomously-triggered antennas, spaced 1 km apart. A large array is necessary to sufficiently sample the radio footprint of an up-going EAS, as they are typically highly-inclined. A total of 20 arrays are planned, and if distributed globally, would provide nearly full sky coverage. By capturing the entire radio footprint, GRAND can distinguish neutrino signals from anthropogenic and cosmic ray backgrounds. The many baselines of a sparse array also provide an excellent pointing resolution.

The Beamforming Elevated Array for COsmic Neutrinos (BEACON) concept consists of multiple compact digitally-phased arrays, placed on high elevation mountains \cite{Wissel_2020}. The high elevation sites allow each array to monitor a large area for up-going EAS, while phasing improves the signal-to-noise ratio of received signals and thus lowers the energy threshold. Together, these characteristics provide BEACON with a very large instantaneous effective area in an efficient manner \cite{Zeolla:2025afb}.

The newly proposed Hybrid Elevated Radio Observatory for Neutrinos (HERON) combines these two concepts, exploiting the advantages of both \cite{HERON}. The concept consists of multiple phased arrays embedded within a large sparse array, distributed along the side of a mountain. Triggering is performed with the phased arrays, while shower reconstruction and background rejection are performed with the sparse array. HERON is designed to target the astrophysical neutrino flux at 100 PeV. It will thus bridge the gap between IceCube and future UHE experiments. Here, we detail the simulations that have been used to optimize the design of HERON. 

\section{Phased Array Simulations}
\subsection{Maximizing Sensitivity at 100 PeV}

To optimize the sensitivity of the phased arrays at 100 PeV, the Monte Carlo MARMOTS \cite{Zeolla:2025afb} has been used. MARMOTS can calculate the instantaneous effective area of any configuration of phased arrays. It does this by calculating the geometric area in view of each array, populating that area with exiting tau-leptons, determining the probability of each exit, and then determining if the resulting phased voltage signal-to-noise ratio exceeds a chosen threshold (nominally 5 $\times$ thermal noise). By varying the design of the phased arrays input into the Monte Carlo, we can thus isolate the variables which improve the sensitivity at 100 PeV.

\begin{figure}[tbp]
\centering
\includegraphics[width=0.5\textwidth]{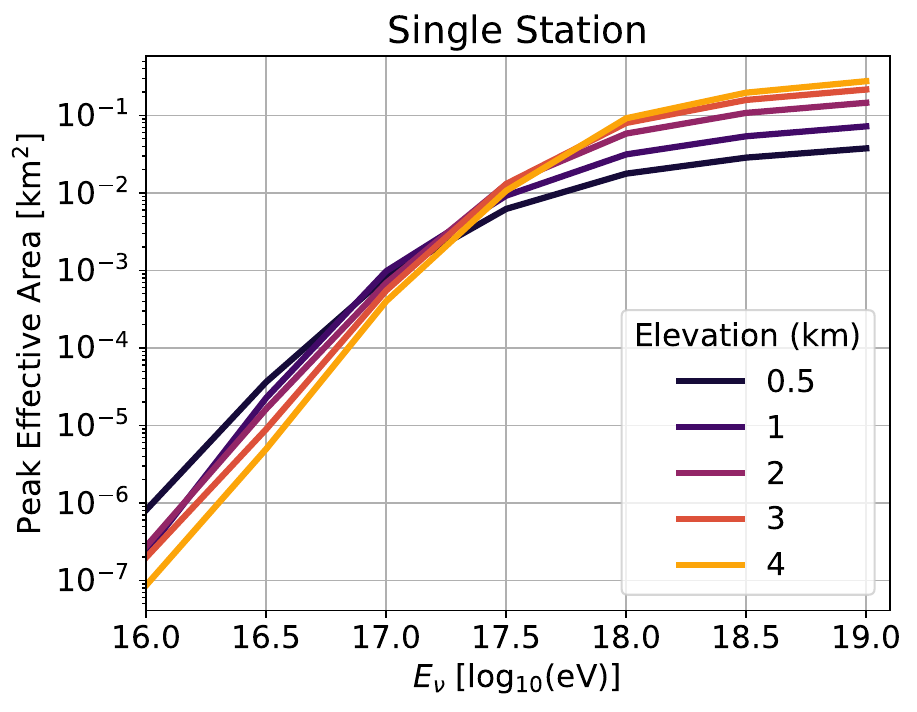}
\caption{Peak effective area as a function of neutrino energy for a single phased array. A range of detector elevations is plotted.}
\label{fig:altitude} 
\end{figure}

One such variable is detector altitude. Placing the phased array at higher altitude increases the geometric area in view, however it also increases the potential distance between decay and detector. The effect of phased array altitude on the peak effective area as a function of neutrino energy is shown in Fig.~\ref{fig:altitude}. We find that below $E_\nu=10^{17.25}$ eV, the negative effect of increasing distance becomes dominant over the positive effect of increasing geometric area. To optimize HERON to 100 PeV, we have therefore targeted an altitude of 1 km. This is considerably lower than the altitude targeted by BEACON of 3 km, improving the effective area at $10^{17}$ eV by a factor of $\sim2$.

The antenna model also has a significant impact on sensitivity. MARMOTS allows the input of antenna simulations, as well as simple models such as isotropic matched antennas with specified gains and bandwidths. Gain has a straightforward relationship with sensitivity, in that higher gain means greater sensitivity. Generally however, higher gain antennas are more directional, shrinking the geometric area observed by each phased array. A balance must therefore be decided between sensitivity and field-of-view.

\begin{figure}[htbp]
 \centering
 \subfloat{\includegraphics[width=0.49\textwidth]{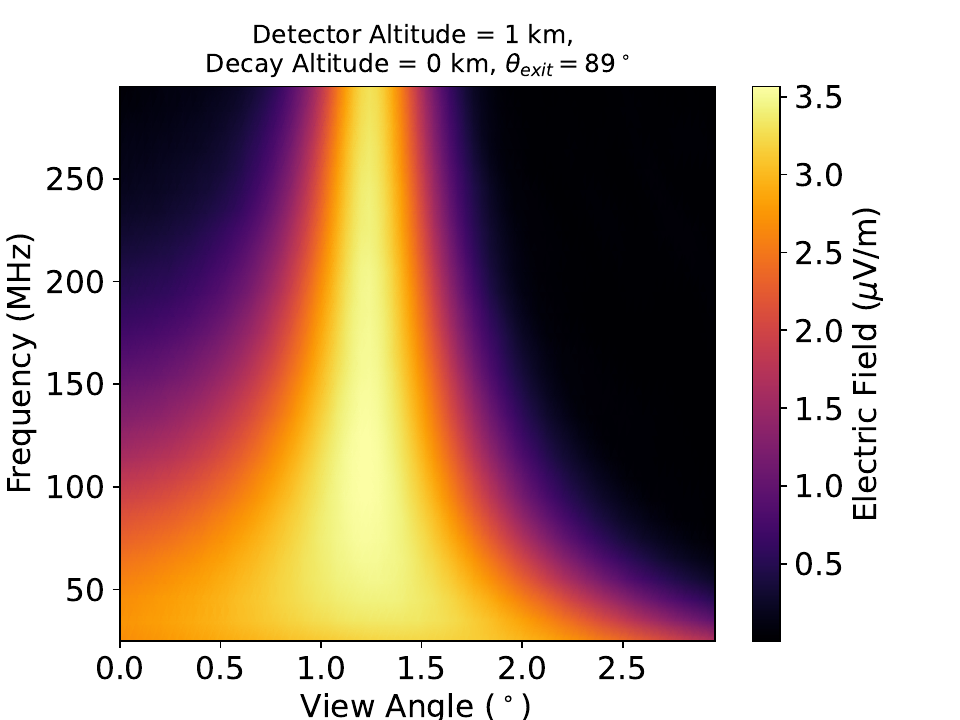}}
 %\,
 \subfloat{\includegraphics[width=0.49\textwidth]{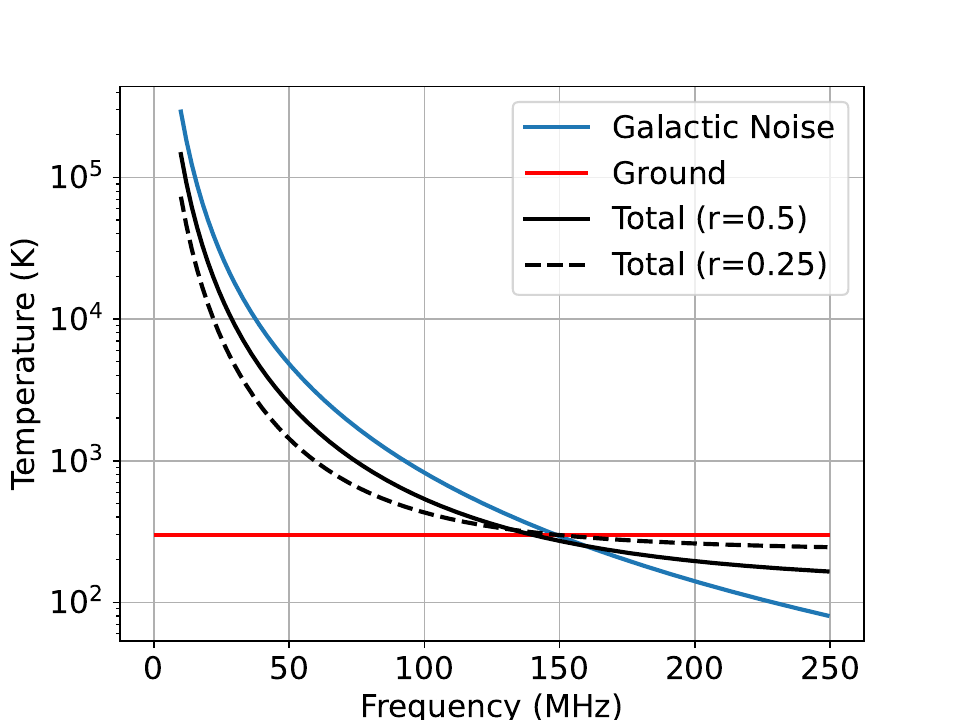}}
  \caption{Left: The peak electric field, bandpass filtered within 10 MHz bins, for a tau-induced EAS as a function of frequency and view angle for a typical Earth-skimming geometry. Simulated using ZHAireS-RASPASS \cite{Tueros:2023mxa,Tueros:2024bzl}. Right: Galactic and ground noise temperature as a function of frequency. Also shown is the total noise temperature for two different sky-fractions. Galactic noise modeled using the Dulk parameterization \cite{Dulk}.}
  \label{fig:freq}
\end{figure}

Bandwidth has a complex relationship with sensitivity due to the relationship between signal, noise, and frequency. Shown on the left in Fig.~\ref{fig:freq} is the peak electric field as a function of frequency and view angle for a typical Earth-skimming tau decay geometry. We see that at low frequencies the electric field is dispersed across a wider range of view angles, allowing detection of EAS from further off axis. At higher frequencies, emission is highly beamed along the Cherenkov angle. On the other hand, shown on the right in Fig.~\ref{fig:freq}, is the noise temperature as a function of frequency. We see that galactic noise rises quickly at lower frequencies. The benefits of targeting the lower frequencies in signal are therefore partially counteracted by the high noise temperature present there. MARMOTS can be used to find the optimal bandwidth for balancing these effects.

Also shown on the right in Fig.~\ref{fig:freq} is the effect of sky fraction ($r$). Sky fraction is defined as the percentage of the antenna's field-of-view which is occupied by the sky. The total noise temperature $T$ is given by $T = r\,T_\text{sky} + (1-r)\, T_\text{ground}$. For an isotropic antenna, $r = 0.5$. Below 150 MHz, where $T_\text{sky}$ is dominant, reducing the sky fraction reduces the total noise. This can be accomplished using a directional antenna, oriented to observe the ground just below the horizon. Sky fraction can be specified in MARMOTS, or calculated from a simulated antenna model.

\subsection{Phasing Efficiency}

In a digitally-phased array, the phased voltage signal is nominally a factor $N$ greater than the signal present in a single antenna, where $N$ is the number of phased antennas. Increasing the number of antennas therefore improves sensitivity. Ideally then, each HERON station would contain the maximum number of antennas that current Field-Programmable Gate Array (FPGA) technology allows. The factor of $N$ improvement, however, assumes that the antennas are sufficiently close together such that their received signals remain similar. If many antennas are included in a single phased array, it is possible that due to the size of the array, the factor of $N$ no longer holds. It is therefore important to discover the maximum distance at which two antennas can be spaced and still maintain phasing efficiency.

\begin{figure}[bp]
    \centering
    \includegraphics[width=0.75\textwidth]{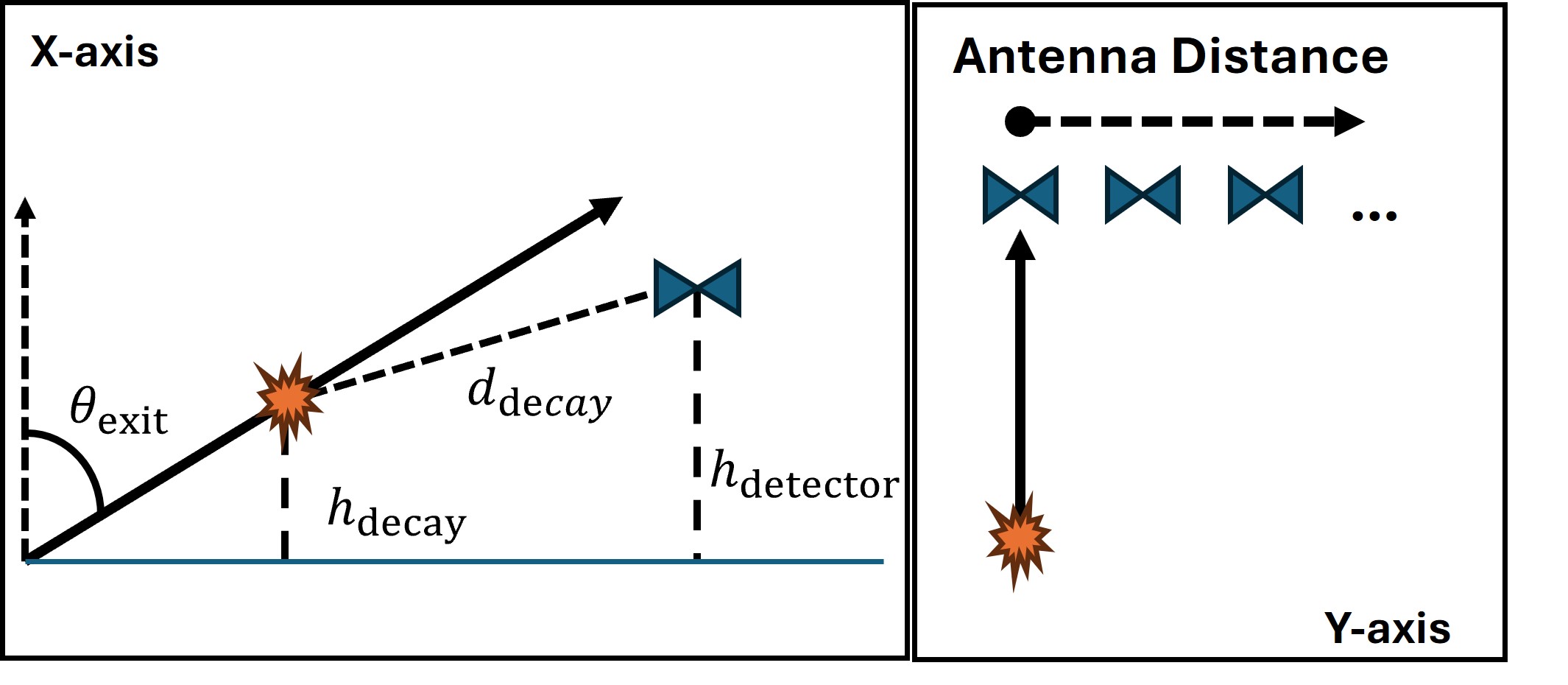}
    \caption{Geometry of the phasing efficiency simulations. Showers are propagated along the x-axis, while antennas are distributed evenly along the y-axis.}
    \label{fig:phasing_geometry}
\end{figure}

\begin{figure}[tbp]
 \centering
 \subfloat{\includegraphics[width=0.49\textwidth]{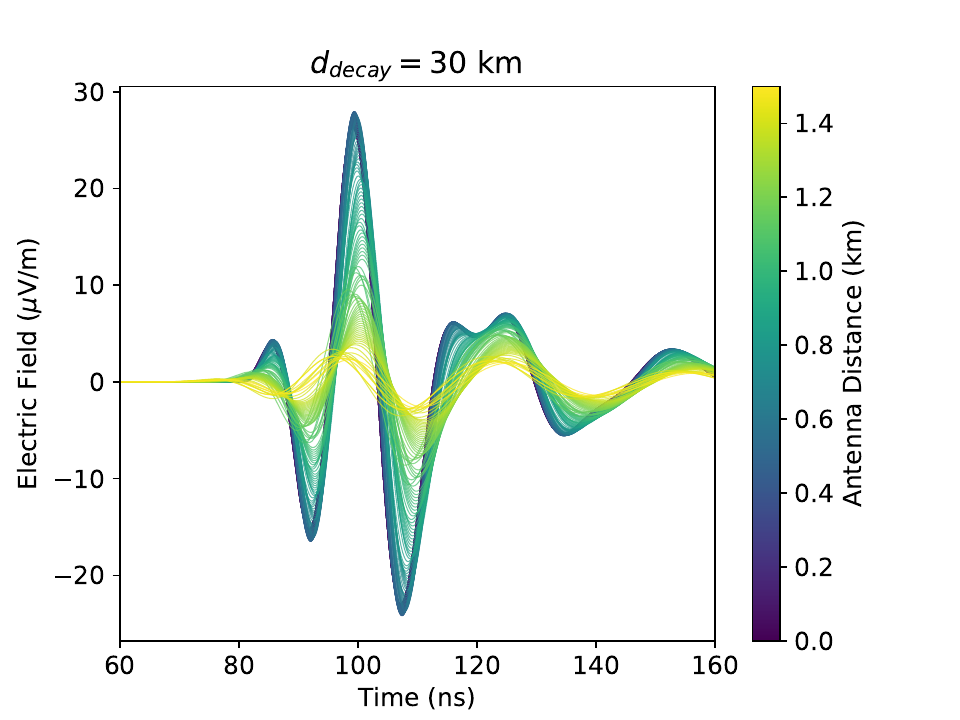}}
 %\,
 \subfloat{\includegraphics[width=0.49\textwidth]{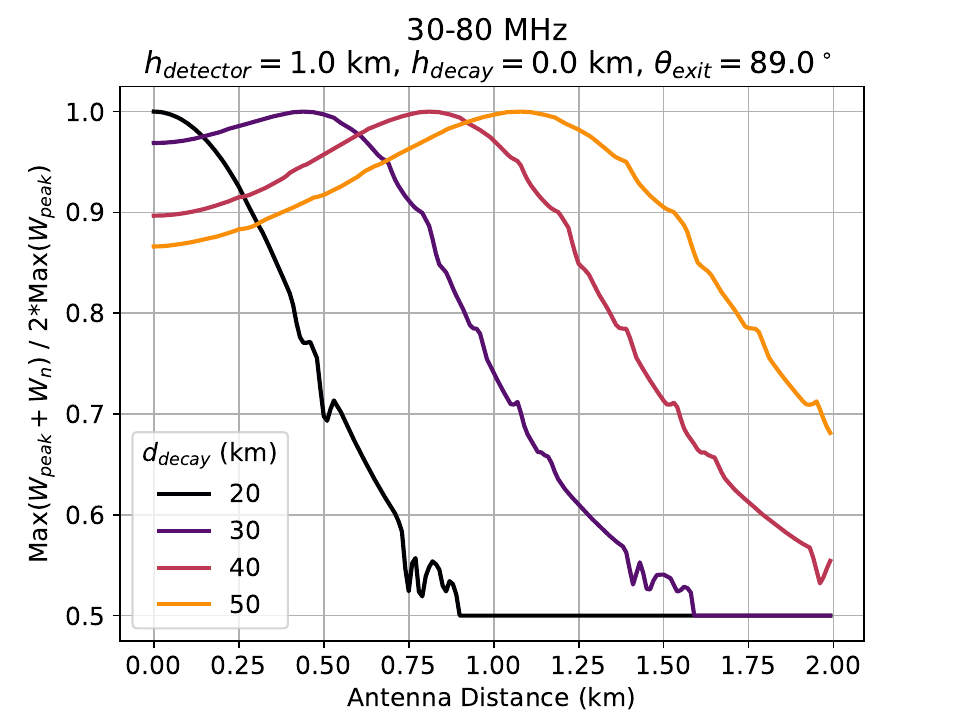}}
  \caption{Left: Example time-domain electric field traces from ZHAireS-RASPASS. Waveforms are colored by the antenna's distance along the $y$-axis. Right: The phasing efficiency as a function of antenna distance for a typical Earth-skimming geometry, for a range of tau decay distances.}
  \label{fig:phasing}
\end{figure}

To this end, a library of ZHAireS-RASPASS \cite{Tueros:2023mxa,Tueros:2024bzl} simulations was generated. In each simulation, the time-domain electric fields resulting from a tau decay with a specified zenith angle ($\theta_\text{exit}$) and decay altitude ($h_\text{decay}$) were calculated for a line of antennas placed at a specified altitude ($h_\text{detector}$). The tau decay propagated along the $x$-axis, while the line of antennas was placed along the $y$-axis. This geometry is depicted schematically in Fig.~\ref{fig:phasing_geometry}. Shown on the left in Fig.~\ref{fig:phasing} are the time-domain electric fields for a typical Earth-skimming geometry ($h_\text{detector}=1$ km, $h_\text{decay}=0$ km, $\theta_\text{exit}=89^\circ$), colored by the antenna's distance from $y=0$. Here, the electric fields have been bandpass filtered to 30-80 MHz, the bandwidth currently employed by the phased array of BEACON.

For sufficiently close together antennas, the sum of waveforms $W_1$ and $W_2$ should have an amplitude about twice that of either $W_1$ or $W_2$. Taking the peak-amplitude waveform, $W_\text{peak}$, and adding it sequentially to each of the other waveforms $W_n$, we can thus define the phasing efficiency as $\text{Max}(W_\text{peak}+W_n) \, / \,[2\cdot\text{Max}(W_\text{peak})]$. For perfect phasing this value will be 1. On the right in Fig.~\ref{fig:phasing}, we show the phasing efficiency as a function of distance along the $y$-axis for a variety of decay distances ($d_\text{decay}$). We find that for all tested decay distances, 90\% phasing efficiency is maintained for antennas within 250 meters of each other. Considering this, and limitations presented by FPGA technology, we infer 24 antennas per phased array is feasible.

\section{Sparse Array Simulations}

\begin{figure}[tbp]
 \centering
 \subfloat{\includegraphics[width=0.5\textwidth]{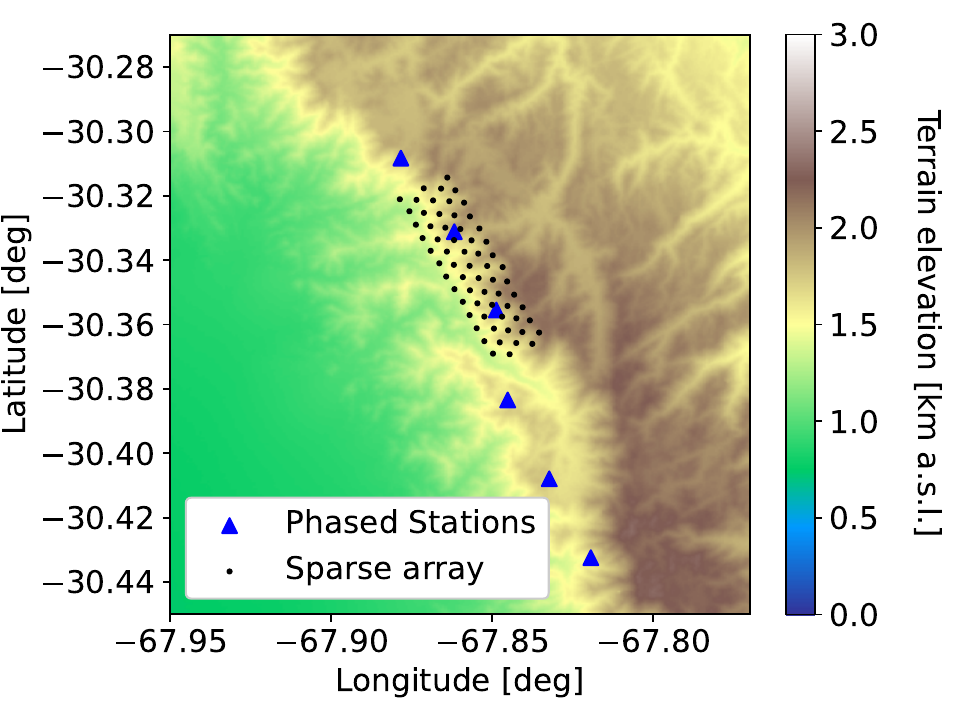}}
 \\
 \subfloat{\includegraphics[width=0.8\textwidth]{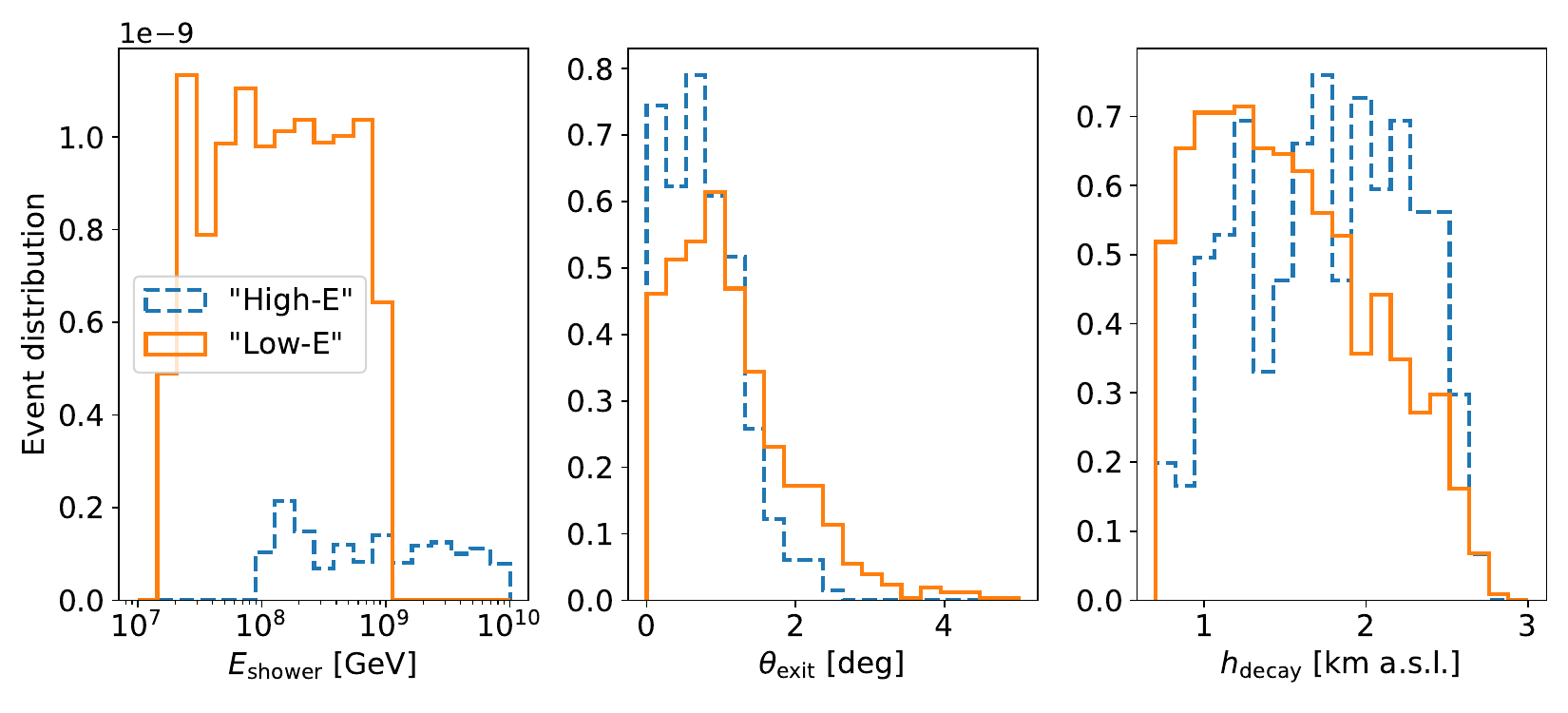}}
  \caption{Top: The simulated sparse array, located in the San Juan province of Argentina. The elevation of each antenna has been determined by the topography. Bottom: the distributions of shower energy, exit angle, and decay altitude resulting from DANTON, for both energy sets.}
  \label{fig:sparse_array}
\end{figure}

A potential experiment site has been identified in the San Juan province of Argentina. To test the interferometric reconstruction capabilities of a sparse array located at this site, a suite of simulations were run for a subset of the total array. The subset consists of a triangular grid of 65 antennas, spaced 500 meters apart. This configuration spans 6 km in width, the planned distance between three phased arrays. The elevation of each antenna was determined from topographic elevation data. A map of this configuration is shown on the left in Fig.~\ref{fig:sparse_array}.

The simulation package DANTON \cite{DANTON} was then used to randomly sample tau decays with off-axis viewing angles less than $3^\circ$ from the center of the array. Two sets were generated: 250 events in the "high-energy" set with $E_\text{shower}\in[10^8-10^{10}]$ GeV, and 972 events in the "low-energy" set with $E_\text{shower}\in [10^{7.3}-10^{9}]$ GeV. The distributions of shower energy, exit angle, and decay altitude for these two sets are shown on the right in Fig.~\ref{fig:sparse_array}. At each antenna, the time-domain electric field generated by the resulting EAS was then simulated using ZHAires-RASPASS.

For a coherent source of radio, emitting a spherical wavefront, the signals will arrive at each antenna with a time delay given by the distance to the source and the speed of light. In radio interferometry, the 3D space around an array is scanned for the potential source location, delaying and summing the waveforms according to the predicted time delays at each point. The point which corresponds to the actual location of the source will result in the signals being aligned, and thus maximum coherence once they are summed. This method can be used to reconstruct the propagation axis of an EAS, as well as to find the location of $X_\text{max}$ \cite{Schoorlemmer:2020low}.

\begin{figure}[tbp]
 \centering
 \subfloat{\includegraphics[width=0.45\textwidth]{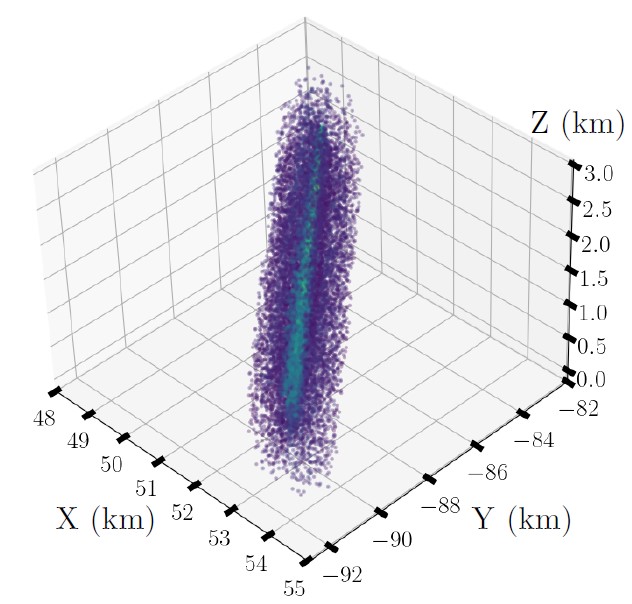}}
 %\,
 \subfloat{\includegraphics[width=0.54\textwidth]{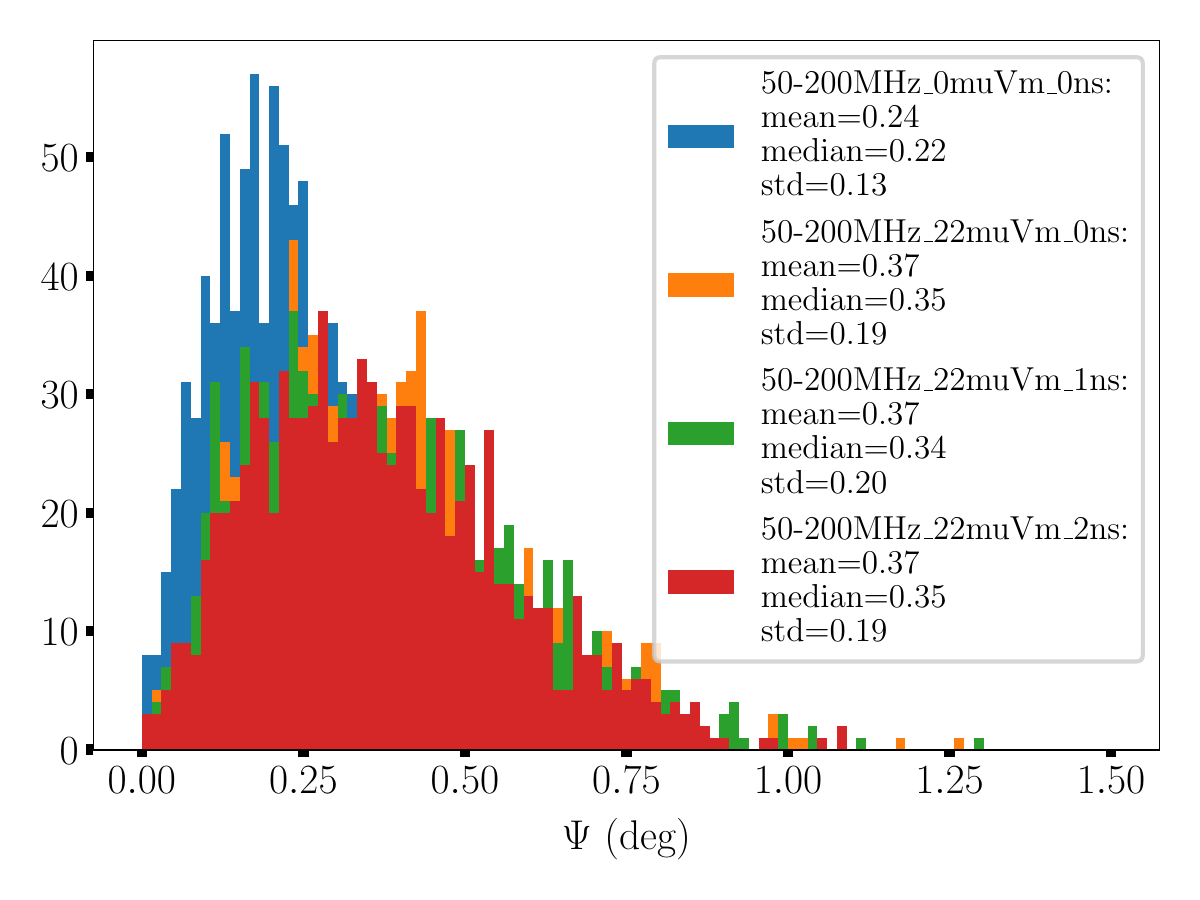}}
  \caption{Left: An example of the interferometric reconstruction technique using a DANTON + RASPASS tau-induced EAS. Potential source locations are sampled from a cylinder around the true particle axis. Each point is colored by the resulting summed signal strength, or coherence. Coherence is greater along the particle axis, and maximized near $X_\text{max}.$ Right: Histograms of the angular accuracy resulting from interferometric reconstruction when using the sparse array simulation suite. Shown in blue is the accuracy with no instrumental effects, in orange the effects of gaussian noise ($\sigma=22$ $\mu$V/m), in green the effects of gaussian noise and a time jitter of 1 ns, and in red the effects of gaussian noise and a time jitter of 2 ns.}
  \label{fig:inter}
\end{figure}

Shown on the left in Fig.~\ref{fig:inter} is an example of this method, in which points were sampled from a cylinder around the true particle axis. Coherence is larger along the particle axis, allowing the axis to be reconstructed via line-fitting. Specifically, the direction is reconstructed by solving for the eigenvector of the covariance matrix with the largest eigenvalue, after weighting the points with a softmax function. For now, the location of maximum coherence is estimated to be $X_\text{max}$. On the right in Fig.~\ref{fig:inter} are histograms of the angular error in reconstruction when applying this method to the DANTON + RASPASS "low-energy" simulation set. Electric fields have been filtered to 50-200 MHz, the bandwidth of the current GRAND sparse array antennas. Shown is the accuracy when 1) there is no noise or time jitter (blue), 2) there is gaussian noise of $\sigma=22$ $\mu$V/m (orange), 3) there is gaussian noise and a time jitter of 1 ns (green), and 4) there is gaussian noise and a time jitter of 2 ns (red). We find that even in the presence of noise and 2 ns time jitter, an average angular resolution of $0.4^\circ$ is achieved. In the future, this method will be applied to the "high-energy" simulation set as well, which will likely achieve even better results. This method will also be tested using waveforms which have been convolved with an antenna and signal-chain response.

\section{Conclusions}

HERON is being optimized to detect astrophysical neutrinos at 100 PeV. It will thus bridge the gap between IceCube and future UHE neutrino detectors. HERON consists of phased arrays, like those employed by BEACON, embedded within a larger sparse array, like those employed by GRAND. Together, this setup enables HERON to exploit the advantages of both designs.

The simulation package MARMOTS is being used to maximize the sensitivity of the phased arrays. In particular, the antenna design seems to play a critical role. Future simulations using MARMOTS will be used to test realistic antenna designs, and select the one which performs best. Using RASPASS simulations, we find that efficient phasing is possible for antennas spaced less than $\sim250$ m apart, enabling relatively large phased arrays. From a suite of DANTON + RASPASS simulations, we find that the sparse array of HERON can achieve a pointing resolution of $0.4^\circ$. Future simulations will incorporate baselines from the phased arrays, further enhancing the pointing resolution.

\bibliographystyle{JHEP}
\bibliography{biblio.bib}

%The following list of authors, affiliations and funding agencies will be updated at the day of submission. The following template is a placeholder generated via https://authorlist.icecube.wisc.edu/icecube on May 17, 2025 and will be updated.
\section*{Full Author List: BEACON Collaboration (July 1st, 2025)}

\scriptsize
\noindent
J.~Alvarez-Mu\~{n}iz$^{1}$,
S.~Cabana-Freire$^{1}$,
W.~Carvalho~Jr.$^{2}$,
A.~Cummings$^{3,4,5}$,
C.~Deaconu$^{6}$,
J.~Hinkel$^{3}$,
K.~Hughes$^{7}$,
R.~Krebs$^{3,4}$,
Y.~Liu$^{7}$,
Z.~Martin$^{8}$,
K.~Mulrey$^{9,10}$
A.~Nozdrina$^{7}$,
E.~Oberla$^{6}$,
S.~Prohira$^{11}$,
A.~Romero-Wolf$^{12}$,
A.~G.~Vieregg$^{6,8,13}$,
S.~A.~Wissel$^{3,4,5}$,
E.~Zas$^{1}$,
A.~Zeolla$^{3,4}$
\\
\\
\noindent
$^{1}$Instituto Galego de F\'\i sica de Altas Enerx\'\i as IGFAE, Univerisade de Santiago de Compostela, 15782 Santiago de Compostela, Spain \\
$^{2}$Faculty of Physics, University of Warsaw, 02-093, Warsaw, Poland \\
$^{3}$Department of Physics, Pennsylvania State University, University Park, PA 16802, USA \\
$^{4}$Center for Multimessenger Astrophysics, Institute of Gravitation and the Cosmos, Pennsylvania State University, University Park, PA 16802, USA \\
$^{5}$Department of Astronomy and Astrophysics, Pennsylvania State University, University Park, PA 16802, USA \\
$^{6}$Department of Astronomy and Astrophysics, Kavli Institute for Cosmological Physics, University of Chicago, Chicago, IL 60637, USA \\
$^{7}$,Department of Physics, The Ohio State University, Columbus, OH 43210, USA \\
$^{8}$Department of Physics, Kavli Institute for Cosmological Physics, University of Chicago, Chicago, IL 60637, USA \\
$^{9}$Department of Astrophysics / IMAPP, Radboud University Nijmegen, 6500 GL, Nijmegen, The Netherlands \\
$^{10}$NIKHEF, Science Park Amsterdam, 1098 XG, Amsterdam, The Netherlands \\
$^{11}$Department of Physics and Astronomy, University of Kansas, Lawrence, KS 66045, USA \\
$^{12}$Jet Propulsion Laboratory, California Institute for Technology, Pasadena, CA 91109, USA \\
$^{13}$Enrico Fermi Institute, University of Chicago, Chicago, IL 60637, USA

\subsection*{Acknowledgments}
\noindent
This work is supported by NSF Awards $\#$ 2033500, 1752922,
1607555, and DGE-1746045 as well as the Sloan Foundation,
the RSCA, the Bill and Linda Frost Fund at the California Polytechnic State University, and NASA (support through JPL and Caltech as well as Award $\#$ 80NSSC18K0231). This work has received financial support from Ministerio de Ciencia, Innovaci\'on y Universidades/Agencia Estatal de Investigaci\'on, MICIU/AEI/10.13039/501100011033, Spain (PID2022-140510NB-I00, PCI2023-145952-2, RYC2019-027017-I, and Mar\'\i a de Maeztu grant CEX2023-001318-M); Xunta de Galicia, Spain (CIGUS Network of Research Centers and Consolidaci\'on 2021 GRC GI-2033 ED431C-2021/22 and 2022 ED431F-2022/15); and Feder Funds. We thank the NSF-funded White Mountain Research Station for their support. Computing resources were provided by the University of Chicago Research Computing Center and the Institute for Computational and Data Sciences at Penn State.

\section*{Full Author List: GRAND Collaboration}

\scriptsize
\noindent
J.~Álvarez-Muñiz$^{1}$, R.~Alves Batista$^{2, 3}$, A.~Benoit-Lévy$^{4}$, T.~Bister$^{5, 6}$, M.~Bohacova$^{7}$, M.~Bustamante$^{8}$, W.~Carvalho$^{9}$, Y.~Chen$^{10, 11}$, L.~Cheng$^{12}$, S.~Chiche$^{13}$
% ...
\\
\\
$^{1}$Departamento de Física de Particulas \& Instituto Galego de Física de Altas Enerxías, Universidad de Santiago de Compostela, 15782 Santiago de Compostela, Spain \\
$^{2}$Institut d'Astrophysique de Paris, CNRS  UMR 7095, Sorbonne Université, 98 bis bd Arago 75014, Paris, France \\
$^{3}$Sorbonne Université, Université Paris Diderot, Sorbonne Paris Cité, CNRS, Laboratoire de Physique 5 Nucléaire et de Hautes Energies (LPNHE), 6 4 place Jussieu, F-75252, Paris Cedex 5, France \\
$^{4}$Université Paris-Saclay, CEA, List,  F-91120 Palaiseau, France \\
$^{5}$Institute for Mathematics, Astrophysics and Particle Physics, Radboud Universiteit, Nijmegen, the Netherlands \\
$^{6}$Nikhef, National Institute for Subatomic Physics, Amsterdam, the Netherlands \\
$^{7}$Institute of Physics of the Czech Academy of Sciences, Na Slovance 1999/2, 182 00 Prague 8, Czechia \\
$^{8}$Niels Bohr International Academy, Niels Bohr Institute, University of Copenhagen, 2100 Copenhagen, Denmark \\
$^{9}$Faculty of Physics, University of Warsaw, Pasteura 5, 02-093 Warsaw, Poland \\
$^{10}$Key Laboratory of Dark Matter and Space Astronomy, Purple Mountain Observatory, Chinese Academy of Sciences, 210023 Nanjing, Jiangsu, China \\
$^{11}$School of Astronomy and Space Science, University of Science and Technology of China, 230026 Hefei Anhui, China \\
$^{12}$National Astronomical Observatories, Chinese Academy of Sciences, Beijing 100101, China \\
$^{13}$Inter-University Institute For High Energies (IIHE), Université libre de Bruxelles (ULB), Boulevard du Triomphe 2, 1050 Brussels, Belgium \\
% ...

%%%%%%%%%%%%%%%%%%%%%%%%%%%%%%%%%%%%%%%%%%%%%%%%%%%%%%%%%%%%%%
%%%%%%%%%%%%%%%%%%%%%%%%%%%%%%%%%%%%%%%%%%%%%%%%%%%%%%%%%%%%%%

\subsection*{Acknowledgments}

\noindent
The GRAND Collaboration is grateful to the local government of Dunhuag during site survey and deployment approval, to Tang Yu for his help on-site at the GRANDProto300 site, and to the Pierre Auger Collaboration, in particular, to the staff in Malarg\"ue, for the warm welcome and continuing support.
The GRAND Collaboration acknowledges the support from the following funding agencies and grants.
%%%%
\textbf{Brazil}: Conselho Nacional de Desenvolvimento Cienti\'ifico e Tecnol\'ogico (CNPq); Funda\c{c}ão de Amparo \`a Pesquisa do Estado de Rio de Janeiro (FAPERJ); Coordena\c{c}ão Aperfei\c{c}oamento de Pessoal de N\'ivel Superior (CAPES).
%%%%
\textbf{China}: National Natural Science Foundation (grant no.~12273114); NAOC, National SKA Program of China (grant no.~2020SKA0110200); Project for Young Scientists in Basic Research of Chinese Academy of Sciences (no.~YSBR-061); Program for Innovative Talents and Entrepreneurs in Jiangsu, and High-end Foreign Expert Introduction Program in China (no.~G2023061006L); China Scholarship Council (no.~202306010363); and special funding from Purple Mountain Observatory.
%%%%
\textbf{Denmark}: Villum Fonden (project no.~29388).
%%%%
\textbf{France}: ``Emergences'' Programme of Sorbonne Universit\'e; France-China Particle Physics Laboratory; Programme National des Hautes Energies of INSU; for IAP---Agence Nationale de la Recherche (``APACHE'' ANR-16-CE31-0001, ``NUTRIG'' ANR-21-CE31-0025, ANR-23-CPJ1-0103-01), CNRS Programme IEA Argentine (``ASTRONU'', 303475), CNRS Programme Blanc MITI (``GRAND'' 2023.1 268448), CNRS Programme AMORCE (``GRAND'' 258540); Fulbright-France Programme; IAP+LPNHE---Programme National des Hautes Energies of CNRS/INSU with INP and IN2P3, co-funded by CEA and CNES; IAP+LPNHE+KIT---NuTRIG project, Agence Nationale de la Recherche (ANR-21-CE31-0025); IAP+VUB: PHC TOURNESOL programme 48705Z. 
%%%%
\textbf{Germany}: NuTRIG project, Deutsche Forschungsgemeinschaft (DFG, Projektnummer 490843803); Helmholtz—OCPC Postdoc-Program.
%%%%
\textbf{Poland}: Polish National Agency for Academic Exchange within Polish Returns Program no.~PPN/PPO/2020/1/00024/U/00001,174; National Science Centre Poland for NCN OPUS grant no.~2022/45/B/ST2/0288.
%%%%
\textbf{USA}: U.S. National Science Foundation under Grant No.~2418730.
%%%
Computer simulations were performed using computing resources at the CCIN2P3 Computing Centre (Lyon/Villeurbanne, France), partnership between CNRS/IN2P3 and CEA/DSM/Irfu, and computing resources supported by the Chinese Academy of Sciences.

\section*{Full Author List: Additional Authors}

\scriptsize
\noindent
\scriptsize
\noindent
I. Allekotte$^{1}$, Luciano Ferreyro$^{2}$, Matias Hampel$^{2}$, F. Sanchez$^{2}$
\\
\\
$^{1}$Centro Atómico Bariloche and Instituto Balseiro (CNEA-UNCuyo-CONICET), San Carlos de Bariloche, Argentina \\
$^{2}$Instituto de Tecnologías en Detección y Astropartículas (CNEA, CONICET, UNSAM), Buenos Aires, Argentina

\end{document}